\documentclass[twocolumn,showpacs,superscriptaddress,amsmath,amssymb]{revtex4}

\usepackage{graphicx}
\usepackage{dcolumn}
\usepackage{bm}

\begin{document}
\title{Classification of the symmetry of photoelectron dichroism broken by light}

\author{Y.~Ishida}
\email[]{ishiday@issp.u-tokyo.ac.jp}
\affiliation{Center for Correlated Electron Systems, Institute for Basic Science, Seoul 08826, Republic of Korea}
\affiliation{ISSP, The University of Tokyo, Kashiwa-no-ha, Kashiwa, Chiba 277-8561, Japan}

\author{D.~Chung}
\affiliation{College of Liberal Studies, Seoul National University, Seoul 08826, Republic of Korea}

\author{J.~Kwon}
\affiliation{Center for Correlated Electron Systems, Institute for Basic Science, Seoul 08826, Republic of Korea}
\affiliation{Department of Physics and Astronomy, Seoul National University, Seoul 08826, Republic of Korea}

\author{Y.~S.~Kim}
\affiliation{Center for Correlated Electron Systems, Institute for Basic Science, Seoul 08826, Republic of Korea}
\affiliation{Department of Physics and Astronomy, Seoul National University, Seoul 08826, Republic of Korea}

\author{S.~Soltani}
\altaffiliation[]{Current address: MAX IV Laboratory, Lund University, PO Box 118, SE-221 00 Lund, Sweden}
\affiliation{Center for Correlated Electron Systems, Institute for Basic Science, Seoul 08826, Republic of Korea}
\affiliation{Department of Physics and Astronomy, Seoul National University, Seoul 08826, Republic of Korea}
\affiliation{Institute of Physics and Applied Physics, Yonsei University, Seoul 03722, Republic of Korea}

\author{Y.~Kobayashi}
\affiliation{ISSP, The University of Tokyo, Kashiwa-no-ha, Kashiwa, Chiba 277-8561, Japan}

\author{A.~J.~Merriam}
\affiliation{Lumeras LLC, 207 McPherson Street, Santa Cruz, California 95060, USA}

\author{L.~Yu}
\affiliation{Beijing National Laboratory for Condensed Matter Physics, IOP, CAS, Beijing 100190, China}

\author{C.~Kim}
\affiliation{Center for Correlated Electron Systems, Institute for Basic Science, Seoul 08826, Republic of Korea}
\affiliation{Department of Physics and Astronomy, Seoul National University, Seoul 08826, Republic of Korea}

\date{\today}

\begin{abstract}
We investigate how the direction of polarized light can affect the dichroism pattern seen in angle-resolved photoemission spectroscopy. To this end, we prepared a sample composed of highly-oriented Bi(111) micro-crystals that macroscopically has infinite rotational and mirror symmetry of the point group $\rm{C}_{\infty\rm{v}}$ and examined whether the dichroism pattern retains the $\rm{C}_{\infty\rm{v}}$ symmetry under the stationary configuration of the light and sample. The direction of the light was imprinted in the pattern. Thereby, we apply group theory and classify the pattern with the configuration of light taken into account. We complete the classification by discussing the cases when the out-of-plane component of the polarization can be neglected, when the incidence angle is either 0$^{\circ}$ or 90$^{\circ}$, when the polarization is either elliptic or linear, and also when the sample is a crystal.
\end{abstract}

\maketitle

\section{Introduction}\label{s1}
\begin{figure}[htb]
\begin{center}
\includegraphics{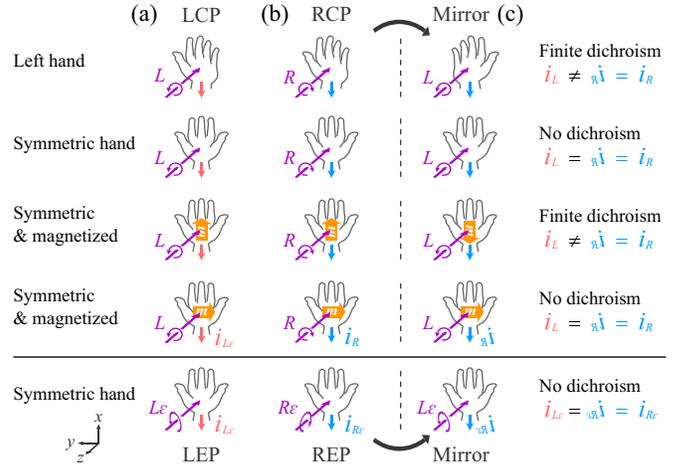}
\caption{\label{fig1} Handedness and dichroism. (a, b) Setups to measure the current $i_{L/L\varepsilon}$ and $i_{R/R\varepsilon}$ when the sample is illuminated with LCP/LEP (a) and RCP/REP (b), respectively. (c) The mirror image of (b). When the images in (a) and (c) are identical, $i_{L/L\varepsilon}=i_{R/R\varepsilon}$. Light is directed along the mirror plane ($y=0$) of the symmetric sample; otherwise, dichroism is allowed to exist in all the setups. ${\bm m}$ is a pseudovector and is invariant with the reflection about $y=0$ when $\bm{m}\parallel$ $y$-axis.}
\end{center}
\end{figure}

The optical response of matter can depend on the polarization of the light. Circular dichroism (CD) is the difference in the response when the polarization is switched from left-circular to right. Similar to the relationship between left and right hands, the left and right circular polarizations (LCP and RCP) are exchanged by a mirror operation. Therefore, the existence of CD can be judged through evaluating the handedness~\cite{90Phys_Schonhense}, or chirality of the experimental setup: Compare the LCP setup and the mirror image of the RCP setup; if the latter can be superimposed on the former, CD does not exist. Some examples of the evaluation after Ref.~\cite{90Phys_Schonhense} are shown in Fig.~\ref{fig1}, in which we included the cases when the samples are magnetized and when the polarizations are elliptic.

In angle-resolved photoemission spectroscopy (ARPES), light illuminates a crystal from a certain direction, and the emitted photoelectrons are analyzed in energy and emission angle. The intensity distribution of the photoelectrons depends on the incident polarization, from which the electronic-state information may be disclosed. Recently, it was articulated that the CD pattern in the distribution could be associated with angular-momentum textures~\cite{11PRL_CD_WangGedik,12PRL_SRPark_Orbital,12NCom_Bahramy_BiSe_2DEG,12PRB_JHHan_OrbitalRashba,12PRB_BYKim_CuAu_OAM,14SciRep_YbB6,15PRL_BiTe_TCChiang,15SciAdv_BiTeX_King,17PRB_Soltani,20PRM_Yilmaz_CrBiSe,20NMat_SWJung_BlackP}, Berry curvatures~\cite{11PRL_Berry_Graphene,18PRL_SCho_Berry}, and degree of surface localization of the wavefunctions~\cite{11PRL_Ishida,17PRB_Kondo}. CD in ARPES~\cite{85PRL_CD_LinearMolecule,89PRL_Schonhense,93JJAP_Daimon,01RepProgPhys_Kuch} has also been utilized in a variety of ways such as to explore the symmetry breaking in cuprate superconductors~\cite{02Nature_Kaminski,04PRL_Borisenko}, to highlight the symmetry-reduced surface states out of bulk states~\cite{07PRB_Vidal_ZnSe}, and to unravel the orbital character of heavy fermions~\cite{08PRL_YbRhSi_Vyalikh}.

While being fruitful, the diverse utility and interpretations of CD ARPES are also under scrutiny~\cite{06PRB_Mulazzi,07PRB_Zabolotnyy,09PRL_Lindroos_TRev_Ortho,09PRB_Mulazzi,11JPCM_Mascaraque_AgGe,12PRL_Bian_BiAg_DivA,13PRL_ZhuDamascielli_Layer,13PRL_RaderScholz_Reversal,13PRB_Madhab_Reversal,13PRB_Vidal_Reversal,13PRB_Arrala_Au111,14PRB_Crepaldi_BiTeX,17PRB_Hanyoung_PhotonE,19PRB_TMD_Ulsrup}. It has been debated which of the symmetry breaking, time reversal or mirror reflection, is sensed in the dichroic signals of the cupartes~\cite{09PRL_Lindroos_TRev_Ortho}; some studies showed that the sign of CD flips when the incident photon energy ($h\nu$) is varied~\cite{09PRB_Mulazzi,13PRL_RaderScholz_Reversal,13PRB_Madhab_Reversal,13PRB_Vidal_Reversal,13PRB_Arrala_Au111}; the CD pattern can also evolve as the incidence angle is increased~\cite{13PRB_Arrala_Au111}. Surprisingly, even the form of the light-electron interaction responsible for photoemission varies among the studies: The starting form is of the dipole interaction~\cite{76SurfSci_Pendry,91PRL_LaanThole}, while some studies include the terms for relativistic correction~\cite{11PRL_CD_WangGedik} and/or surface photoemission~\cite{07PRB_Zabolotnyy,12PRL_Bian_BiAg_DivA}. Thus, it is still under discussion to what extent the magnitude and pattern of the dichroic signals are reflecting the electronic-state properties. A solid basis for understanding CD ARPES is called for.

\begin{figure}[htb]
\begin{center}
\includegraphics{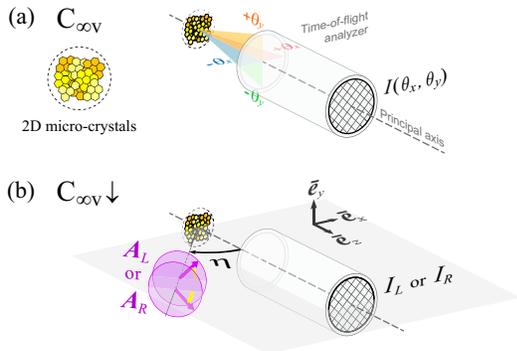}
\caption{\label{fig2} CD ARPES setup in which only light breaks the symmetry of $\rm{C}_{\rm{\infty v}}$. (a) Sample and slit-less photoelectron analyzer set in the $\rm{C}_{\rm{\infty v}}$ configuration. The sample is composed of Bi(111) micro-crystals randomly oriented on HOPG. (b) LCP and RCP incident on the sample-and-analyzer setup. The symmetry is lowered from $\rm{C}_{\rm{\infty v}}$ because of the incidence.}
\end{center}
\end{figure}

In the present study, we focus on the configuration of light that can affect the dichroism pattern in the photoelectron distribution. Significant though it may be, it is not trivial to distinguish the effect of the orientation of light from the others related to the electronic states and light-electron interaction responsible for photoemission. To this end, we performed CD ARPES on a highly symmetrical setup, in which the entity of the sample and electron analyzer has infinite rotational and mirror symmetry characterized by the point group $\rm{C}_{\rm{\infty v}}$ [Fig.~\ref{fig2}(a)]. We let only the incidence of light to break the $\rm{C}_{\rm{\infty v}}$ symmetry of the entire experimental setup [Fig.~\ref{fig2}(b)]. In this way, the effects related to the crystal symmetry can be averaged out and the effect of the orientation of light can be singled out. Namely, we conducted an experiment to clarify whether the light-induced reduction of the symmetry is reflected in the pattern. After the clarification, we present a systematic classification of the pattern aided by the group theory~\cite{76GroupT}. The classification takes into account the configuration of light (incidence angle and ellipticity) and includes the cases when the crystalline sample has some mirror planes.

The paper is structured as follows: After the present Introduction (Section~\ref{s1}), we describe the setup to attain the $\rm{C}_{\rm{\infty v}}$ symmetry of the sample-and-detector entity in Section~\ref{s2}; then, we present the CD ARPES results in Section~\ref{s3}, in which we show that the light-induced reduction of the symmetry is imprinted in the CD pattern; thereby in Section~\ref{s4}, we present a systematic classification of the CD pattern, in which we investigate the cases for a variety of incidence angles, ellipticity, and also when the sample is a crystal; the summary and remarks are made in Section~\ref{s5}; in Appendix~\ref{A1}, we provide the irreducible representations for $\rm{C}_{\rm{\infty v}}$, which is relevant when the light is in normal incidence to the sample surface.

\section{The \textbf{C}$_{\infty\textbf{v}}$ configuration}\label{s2}

The experimental setup is unique in that the entity of the sample and electron analyzer has the symmetry of $\rm{C}_{\rm{\infty v}}$ during the data acquisition. The key to the high-symmetry setup is twofold: (1) We prepared a sample that is effectively $\rm{C}_{\rm{\infty v}}$; (2) we used a slit-less photoelectron analyzer~\cite{88RSI_Daimon,08ApplPhysA_Kirchmann,16UltraMic_Tusche,18RepProgPhys_Zhou,18JAP_Widdra,20JJAP_Matsui}. 

The sample with $\rm{C}_{\rm{\infty v}}$ symmetry was prepared by evaporating bismuth (Bi) of $\sim$100-nm thickness on highly-oriented pyrolytic graphite (HOPG) at $\sim$10$^{-9}$~Torr at room temperature and then heated to 370~K, as described elsewhere~\cite{16RSI_Ishida}. HOPG is composed of stacked layers of graphite micro-crystals oriented randomly in plane, and Bi grown thereon forms into micro-crystals with the (111) face oriented normal to the surface~\cite{16RSI_Ishida,10SurfSci_Brown_AnnTemp,11SurfSci_Brown_STM_XPS,14RSI_Ishida_TRPES}; see the illustration of the sample in Fig.~\ref{fig2}(a). Macroscopically, Bi/HOPG is invariant with respect to whatever rotation about the $z$-axis along the surface normal and whatever mirror reflection about the plane containing the $z$-axis, and thus its symmetry can be characterized by the point group $\rm{C}_{\rm{\infty v}}$.

Photoelectrons were collected by using a slit-less analyzer, namely, the angle-resolved time-of-flight (ARToF) analyzer of Scienta-Omicron. In contrast to the analyzers that can accept photoelectrons emitted into a zero-dimensional hole or one-dimensional slit, the slit-less type can collect photoelectrons emitted into a two-dimensional solid-angular cone. This enabled us to acquire the photoelectron distribution in the cone without rotating the sample, or with the configuration of the light and sample fixed in space. We set the sample surface normal along the principal axis of the ARToF analyzer as shown in Fig.~\ref{fig2}(a), and hence, the symmetry of the entity of the sample and analyzer can be characterized with $\rm{C}_{\rm{\infty v}}$.

\begin{figure*}[htb]
\begin{center}
\includegraphics{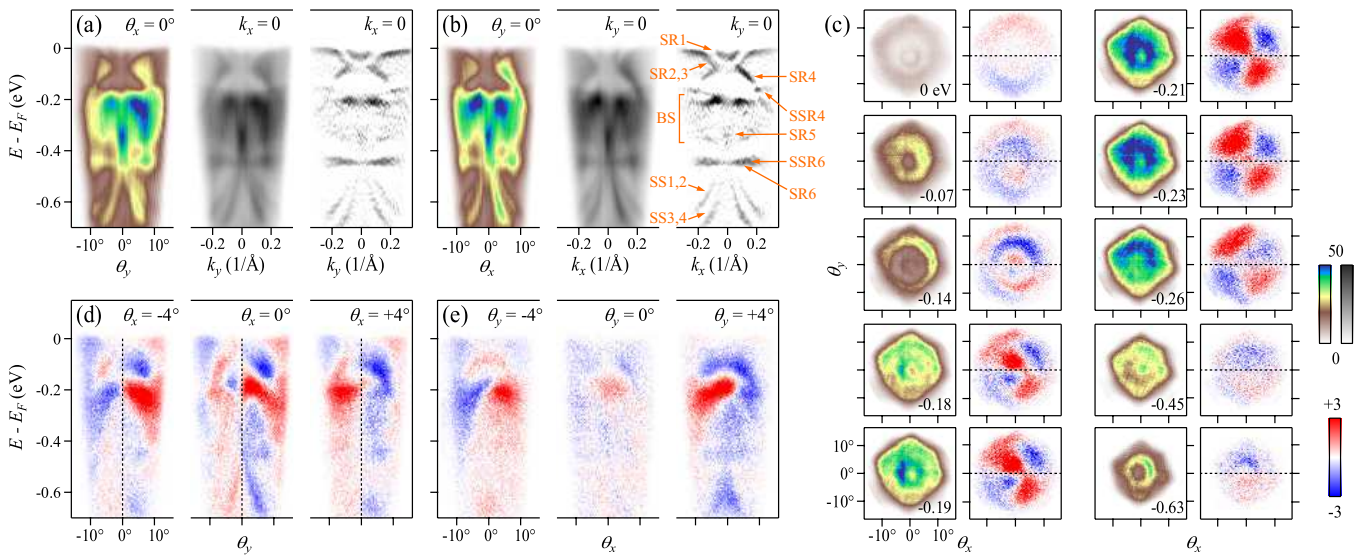}
\caption{\label{fig3}CD ARPES of Bi(111)/HOPG. (a and b) Cuts of the total intensity $I_L+I_R$ at $\theta_x=0^{\circ}$ ($k_x=0$) (a) and at $\theta_y=0^{\circ}$ ($k_y=0$) (b). The left panel is the distribution in energy ($E-E_F$) and emission angle; the middle and right panels are respectively the plots in momentum ($k$) space and its second derivative along the energy. (c) Constant-energy cuts of $I_L+I_R$ and $I_L-I_R$. (d and e) CD seen in the cuts at $\theta_x$ (d) and $\theta_y$ (e). The cuts shown in (a, b, d and e) and (c) are the average of the intensity distributed within $\theta_{x, y}\pm0.5^{\circ}$, and $(E-E_F)\pm3$~meV, respectively.}
\end{center}
\end{figure*}

For the polarized light source, we adopted the 10.8-eV laser-based source~\cite{16RSI_11eV} commercialized by Lumeras. The 10.8-eV harmonics was generated in a xenon-filled tube through a four-wave mixing ($4\varOmega\times2+\varOmega=9\varOmega$) at the repetition rate of 1~MHz, and the polarization of the 10.8-eV output was controlled by varying the polarization of the fundamental laser $\varOmega$~\cite{13OSA_Andrew}. We set the basis of the Cartesian coordinate $\{\bm{\bar e}_x,\bm{\bar e}_y,\bm{\bar e}_z\}$ so that the plane of incidence is $y=0$, and the incidence angle was set to $\eta=50^{\circ}$; see Fig.~\ref{fig2}(b). Because the 10.8-eV beam passed through a Brewster prism after its generation~\cite{16RSI_11eV}, the $\bm{\bar e}_y$ component of the polarization was slightly lost and became elliptic. The effect of the ellipticity will be discussed in Section~\ref{s4d}. The 10.8-eV ARToF system was operated according to Ref.~\cite{18RepProgPhys_Zhou}. The direction of the photoelectron is described with the polar-angular notation $(\theta_x, \theta_y)$~\cite{18RSI_Ishida_Slitless}, where $\theta_x=0^{\circ}$ ($\theta_y=0^{\circ}$) is for the emission into $x=0$ ($y=0$) plane. The temperature of the sample was maintained at $\sim$80~K with liquid nitrogen and the vacuum level of the analyzer chamber was $2\times10^{-10}$~Torr.

\section{Results}\label{s3}

The left panels of Figs.~\ref{fig3}(a) and \ref{fig3}(b) show the cuts of the total photoelectron distribution $I_L+I_R$ at $\theta_x=0^{\circ}$ and $\theta_y=0^{\circ}$, respectively. The middle and right panels are, respectively, the distribution mapped on energy-momentum space and its second derivative along the energy; the latter is the distribution of the negative-curvature strength that highlights the bands. The bands seen in Figs.~\ref{fig3}(a) and \ref{fig3}(b) appear identical. This supports that Bi(111)/HOPG was successfully formed to have the rotational symmetry about the surface normal. Based on the literature~\cite{15JESRP_Hirahara,19JCPS_Bian}, we can assign the bands to surface state (SS), surface resonance (SR), strongly hybridizing surface resonance (SSR), and bulk state (BS), as indicated in Fig.~\ref{fig3}(b). For example, the surface Rashba bands along $\bar{\varGamma}$\,-\,$\bar{M}$ (SR2, 3) and $\bar{\varGamma}$\,-\,$\bar{K}$ (SR4) are simultaneously observed in the cuts. In the panels of Fig.~\ref{fig3}(c), we show the cuts of $I_L+I_R$ at some selected energies. The circular contours seen in the constant-energy cuts further support that the sample was formed into the $\rm{C}_{\rm{\infty v}}$ symmetry. For the cuts of $I_L\pm I_R$ at a variety of energies and angles, see Supplemental Material movie file~\cite{SOM}.

In Fig.~\ref{fig3}(c), we also displayed the constant-energy cuts of the CD distribution $I_L-I_R$. First of all, CD is finite; in other words, CD is not vanished even when the sample-and-analyzer is configured to have the $\rm{C}_{\rm{\infty v}}$ symmetry. 

Patterns in the CD distribution can be characterized by nodes, or where the CD disappears. In the CD patterns of the constant-energy cuts shown in Fig.~\ref{fig3}(c), there is always a horizontal node at $\theta_y=0^{\circ}$ and the pattern appears as a reflection of itself with a sign flip with respect to the node. The magnitude of CD does not strictly obey the anti-symmetry argument partly because the efficiency of the multi-channel-plate detector as well as the transmission of ARToF can be inhomogeneous over the two-dimensional detection plane. Nevertheless, the locus of the node is indifferent to the inhomogeneity of the two-dimensional detection, and therefore, the existence of the horizontal node at $\theta_y=0^{\circ}$ is solid. 

Figures~\ref{fig3}(d) and \ref{fig3}(e) respectively show the cuts of the CD distribution $I_L-I_R$ at some selected $\theta_x$ and $\theta_y$. The signal of CD is substantial in all the cuts except for that at $\theta_y=0^{\circ}$ where the intensity is relatively, or vanishingly, small; this cut corresponds to the horizontal node. The cuts at constant $\theta_x$ [Fig.~\ref{fig3}(d)] are virtually anti-symmetric with respect to the node at $\theta_y=0^{\circ}$. On the other hand, the cuts at $\theta_y$ away from 0$^{\circ}$ [Fig.~\ref{fig3}(e)] exhibit a variety of nodes that winds in the image.   

Summarizing the results, the prepared Bi(111)/HOPG sample successfully formed to have the $\rm{C}_{\rm{\infty v}}$ symmetry, as judged from the patterns of $I_L+I_R$. The CD distribution of the photoelectrons emitted from the $\rm{C}_{\rm{\infty v}}$-symmetric Bi(111)/HOPG was not null but finite, and the pattern of the CD distribution was anti-symmetric with respect to the node on the $y=0$ plane set by the direction of the light.

\section{Discussion}\label{s4}
\subsection{Classifying CD with a point group}\label{s4a}

Taking the opportunity that the setup is highly symmetric, let us apply group theory and interpret the results. It will be shown that the CD distribution $I_L-I_R$ and total distribution $I_L+I_R$ can be related to, if not identified to, the base for representing the symmetry. Hereafter, the incidence angle $\eta$ is neither 0$^{\circ}$ nor 90$^{\circ}$ unless described otherwise. 

\begin{figure}[htb]
\begin{center}
\includegraphics{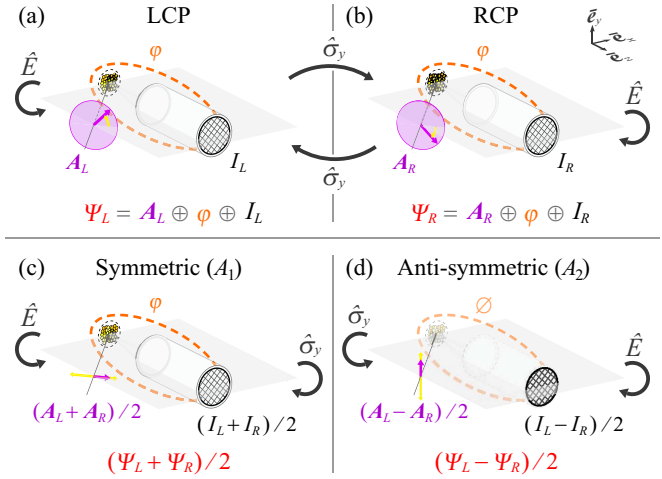}
\caption{\label{fig4} Symmetry operations on the experimental setup. (a) Setup for LCP, $\varPsi_L$. (b) Mirror reflection of $\varPsi_L$, which is identical to the setup for RCP. (c and d) Symmetric (c) and anti-symmetric component (d) extracted from the two illustrations $\varPsi_L$ and $\varPsi_R$. Note, the two no more correspond to any realistic setups but are mathematical entities; the horizontally-polarized-light setup (c) would not result in the distribution $(I_L + I_R)/2$, and vertically-polarized light incident on null sample-and-analyzer (d) is not a realistic experimental setup.}
\end{center}
\end{figure}

Let us regard the illustration of the LCP experiment [Fig.~\ref{fig4}(a)] as a function $\varPsi_L$ in $xyz$ space. The illustration consists of the followings: the sample-and-analyzer $\varphi(x, y, z)$; the polarized photon field ${\bm A}_{L}(x, y, z; \eta)=A({\bm a}_{+}e^{-i\omega t} + c.~c.)$ in the scalar gauge~\cite{76PRB_Kliewer}, where ${\bm a}_{\pm}=(\cos\eta\,{\bm{\bar e}}_x \pm i {\bm{\bar e}}_y + \sin\eta\,{\bm{\bar e}}_z)/2$ is proportional to the polarization vector; and the photoelectron distribution at the detector $I_{L}(x, y, z)$, which is a functional of the polarized photon field. $\varPsi_L$ can be formulated as the direct summation of light, sample-and-analyzer, and the distribution:
\begin{equation}
\varPsi_{L}={\bm A}_{L}\oplus\varphi\oplus I_{L}. \label{eq_directsum}
\end{equation}

Having regarded the illustration as a function, we can now apply operators on it and obtain new illustrations. Here, we apply two operators that consist the point group ${\rm C}_{\sigma}=\{\hat{E}, \hat{\sigma}_y\}$. By applying the mirror operator $\hat{\sigma}_y$ on $\varPsi_L$, a new illustration $\varPsi_{R}$ is constructed: 
\begin{equation}
\varPsi_{R}\equiv\hat{\sigma}_y\varPsi_{L}=\hat{\sigma}_y{\bm A}_{L}\oplus\hat{\sigma}_y\varphi\oplus\hat{\sigma}_yI_{L}={\bm A}_{R}\oplus\varphi\oplus I_{R}; \nonumber
\end{equation}
also see, how the illustration $\varPsi_{R}$ of Fig.~\ref{fig4}(b) is constructed from $\varPsi_{L}$ of Fig.~\ref{fig4}(a). $\varPsi_{R}$ happens to be identical to the illustration of the CD ARPES setup with the incidence of RCP, and therefore, the mirror image of $I_L$ is nothing but the distribution obtained with RCP: $I_R=\hat{\sigma}_yI_L$. Note, $\hat{\sigma}_y{\bm A}_{L}={\bm A}_{-}e^{-i\omega t}+c.\,c.\,={\bm A}_{R}$ because $\hat{\sigma}_y \{\bm{\bar e}_x, \bm{\bar e}_y, \bm{\bar e}_z\}=\{\bm{\bar e}_x, -\bm{\bar e}_y, \bm{\bar e}_z\}$, and $\hat{\sigma}_y\varphi=\varphi$~\cite{footnote1}.

The two illustrations, or the two functions $\varPsi_{L}$ and $\varPsi_{R}$, are exchanged when operated on by $\hat{\sigma}_y$, while they remain themselves when operated on by the identity $\hat{E}$; see Figs.~\ref{fig4}(a) and \ref{fig4}(b). Thus, the set $\{\varPsi_L, \varPsi_R\}$ forms a two-dimensional basis for representing the point group ${\rm C}_{\sigma}$:
\begin{eqnarray}
\hat{E} \{\varPsi_L, \varPsi_R\}	&=& \{\varPsi_L, \varPsi_R\}	\left[
\begin{array}{cc}
1	& 0\\
0	& 1
\end{array}
\right], \label{eq_mat1} \\
\hat{\sigma}_y \{\varPsi_L, \varPsi_R\}	&=& \{\varPsi_L, \varPsi_R\}	\left[
\begin{array}{cc}
0	& 1\\
1	& 0
\end{array}
\right].	\label{eq_mat2}
\end{eqnarray}

\begin{table}
\caption{\label{tab1}Character table for ${\rm C}_{\sigma}$.}
\begin{ruledtabular}
\begin{tabular}{ccrr}
${\rm C}_{\sigma}$&
\textrm{Basis}&
$\hat{E}$&
$\hat{\sigma}_y$\\
\colrule
$A_1$ & $\varPsi_L+\varPsi_R$ & 1 & 1\\
$A_2$ & $\varPsi_L-\varPsi_R$ & 1 & -1\\
\end{tabular}
\end{ruledtabular}
\end{table}

The matrix representation displayed in eqs.~(\ref{eq_mat1}) and (\ref{eq_mat2}) is reducible. It is $\{\varPsi_L+\varPsi_R\}$ and $\{\varPsi_L-\varPsi_R\}$ that respectively form the one-dimensional irreducible representations $A_1$ and $A_2$ of ${\rm C}_\sigma$; see, Table~\ref{tab1}. In terms of the illustration, $(\varPsi_L+\varPsi_R)/2$ and $(\varPsi_L-\varPsi_R)/2$ are the symmetric and anti-symmetric components extracted from the two illustrations $\varPsi_L$ and $\varPsi_R$;  see Figs.~\ref{fig4}(c) and \ref{fig4}(d). $I_L+I_R$ and $I_L-I_R$ are the subset of, or part of the illustration of, $\{\varPsi_L+\varPsi_R\}$ and $\{\varPsi_L-\varPsi_R\}$, respectively. 

To summarize, $I_L+I_R$ and $I_L-I_R$ are respectively identified to {\it the subset of} the bases for the $A_1$ and $A_2$ representations of ${\rm C}_{\sigma}$. If there is no confusion about $I_L\pm I_R$ being the subset of $\{\varPsi_L\pm\varPsi_R\}$, it may be restated as follows: $I_L+I_R$ and $I_L-I_R$ respectively have $A_1$ and $A_2$ symmetry of ${\rm C}_{\sigma}$.

If the photoelectron distribution is irrelevant to the symmetry reduction due to light, then either the illustration of the light is formally omitted from Fig.~\ref{fig4} or the incidence angle $\eta$ is set to zero, and $\{\varPsi_L\pm\varPsi_R\}$ becomes the base for the irreducible representation $D_0^\pm$ of ${\rm C}_{\infty{\rm v}}$; see Appendix~\ref{A1}.  As a result, $I_L-I_R$, which is the subset of  $\{\varPsi_L-\varPsi_R\}$, becomes a null function.

Now, the question asked in Introduction (Section~\ref{s1}) can be reformatted as follows: $I_L-I_R$ is related to the base for a point-group representation, but of which group, ${\rm C}_{\infty{\rm v}}$ or ${\rm C}_{\sigma}$? Given the results shown in Fig.~\ref{fig3}, the answer is ${\rm C}_{\sigma}$, because $I_L-I_R$ is not null but anti-symmetric with respect to the horizontal node.

\subsection{The ${\rm C}_{2\rm{v}}$ case}\label{s4b}
In the previous Section~\ref{s4a}, we showed that $I_L-I_R$ acquires a horizontal node that reflects the symmetry of ${\rm C}_{\sigma}$ even though the sample-and-detector had the ${\rm C}_{\infty{\rm v}}$ symmetry. In this section, we show that there are two cases when $I_L-I_R$ retains a higher symmetry of ${\rm C}_{2\rm{v}}$. We also derive some implications from the fact that $I_L-I_R$ of Bi/HOPG {\it did not} exhibit the pattern of ${\rm C}_{2\rm{v}}$.

\begin{figure}[htb]
\begin{center}
\includegraphics{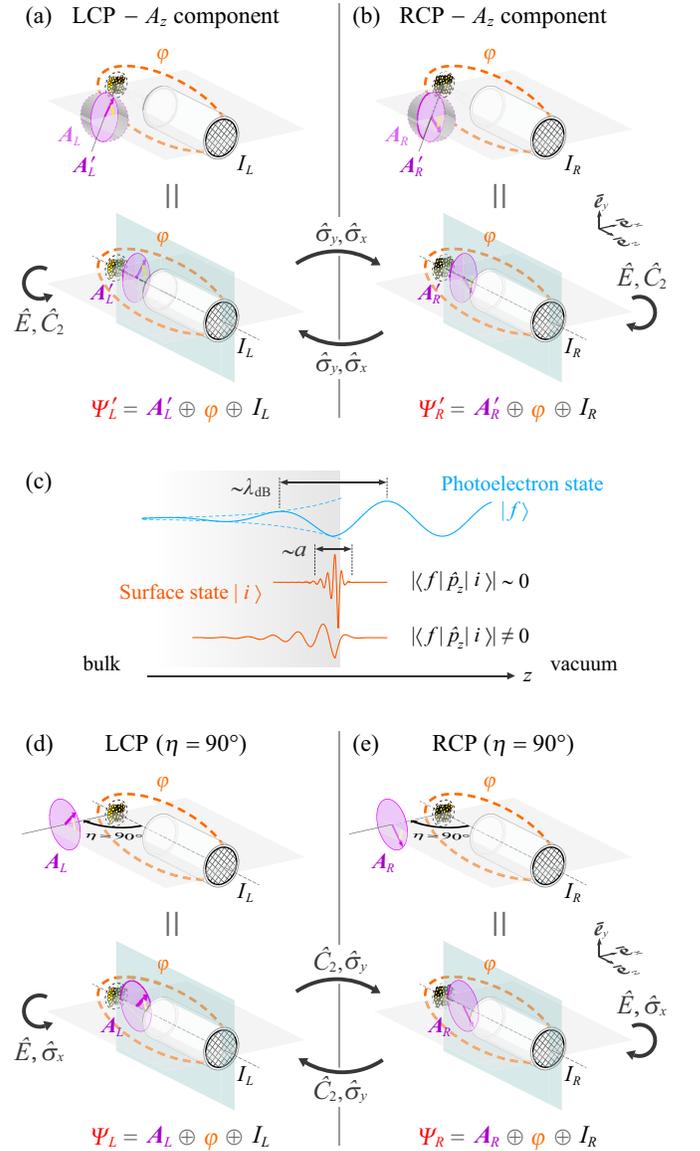}
\caption{\label{fig5}The ${\rm C}_{2{\rm v}}$ case. (a and b) Illustrations for LCP (a) and RCP (b) experiments when the $A_z$ component can be omitted from the light-electron interaction. (c) Spatial profile of the initial and final states. When the initial state is localized on the surface within a length scale $a$ shorter than the de Broglie wavelength of the photoelectron state, then $|\langle f|\hat{p}_z|i\rangle|\sim 0$~\cite{11PRL_Ishida}. (d and e) Illustrations for LCP (d) and RCP (e) experiments when the incidence angle is 90$^{\circ}$.}
\end{center}
\end{figure}

The first case can occur when the $z$ component of the photon field ($A_z={\bm A}\cdot\bm{\bar{e}}_z$) is neglected from the light-electron interaction. As shown in Figs.~\ref{fig5}(a) and \ref{fig5}(b), the $A_z$-omitted photon field ${\bm A}_{L,R}^{'}={\bm A}_{L,R}-({\bm A}_{L,R}\cdot\bm{\bar{e}}_z)\bm{\bar{e}}_z$ orbits on an oval elongated along $y$-axis in the $xy$ plane. The two functions $\varPsi_{L,R}^{'}={\bm A}_{L,R}^{'}\oplus\varphi\oplus I_{L,R}$ are exchanged (remain themselves) when operated on with $\hat{\sigma}_y$ and $\hat{\sigma}_x$ ($\hat{E}$ and $\hat{C}_2$); in other words, the set $\{\varPsi_L^{'},\varPsi_R^{'}\}$ forms a two-dimensional base that represents the point group ${\rm C}_{2{\rm v}}=\{\hat{E},\hat{C}_2,\hat{\sigma}_y,\hat{\sigma}_x\}$. Here, $\hat{C}_2$ and $\hat{\sigma}_x$ are the operators for the two-fold rotation about the principal ($z$) axis and mirror reflection with respect to $x=0$, respectively. It can easily be verified that $\{\varPsi_L^{'}\pm\varPsi_R^{'}\}$ forms the base for the irreducible representation of ${\rm C}_{2{\rm v}}$ (Table~\ref{tab2}). Particularly, $I_L-I_R$ becomes the subset of the base $\varPsi_L^{'}-\varPsi_R^{'}$ for the $A_2$ representation of ${\rm C}_{2{\rm v}}$. Thus, $I_L-I_R$ acquires the vertical node in addition to the horizontal node.

\begin{table}
\caption{\label{tab2} Character table for ${\rm C}_{2{\rm v}}$.}
\begin{ruledtabular}
\begin{tabular}{ccrrrr}
${\rm C}_{2{\rm v}}$ & \textrm{Basis} & $\hat{E}$ & $\hat{C}_2$ & $\hat{\sigma}_y$ & $\hat{\sigma}_x$\\
\colrule
$A_1$ & $\varPsi_L^{'} + \varPsi_R^{'}$, $\varPsi_L + \varPsi_R (\eta=90^{\circ})$ & 1 & 1 & 1 & 1\\
$A_2$ & $\varPsi_L^{'} - \varPsi_R^{'}$ & 1 & 1 & -1 & -1\\
$B_1$ &  & 1 & -1 & 1 & -1\\
$B_2$ &  $\varPsi_L - \varPsi_R (\eta=90^{\circ})$ & 1 & -1 & -1 & 1\\
\end{tabular}
\end{ruledtabular}
\end{table}

Under what condition can $A_z$ be neglected from the light-electron interaction responsible for photoemission? When the dipole approximation is valid so that the surface photoelectric term $\nabla\cdot{\bm A}$ can be neglected, the matrix element of the interaction reads
\begin{equation}
 \langle f|\bm{A}\cdot\hat{\bm{p}}|i\rangle=A_x\langle f|\hat{p}_x|i\rangle+A_y\langle f|\hat{p}_y|i\rangle+A_z\langle f|\hat{p}_z|i\rangle, \label{eq_dipole}
\end{equation}
where $|i\rangle$ and $|f\rangle$ are the initial state and photoelectron final state, respectively. If
\begin{equation}
|\langle f|\hat{p}_{x, y}|i\rangle|\gg|\langle f|\hat{p}_z|i\rangle|\sim 0, \label{cond1}
\end{equation}
then the third term in eq.~\eqref{eq_dipole} and hence $A_z$ can be neglected and the vertical node may emerge. 

Condition~\eqref{cond1} can be fulfilled when the wavefunction of the initial state $\psi_i=\langle xyz|i\rangle$ is localized in the surface region within a length scale $a$ shorter than the de Broglie wavelength $\lambda_{\rm{dB}}$ of the final-state wave function $\psi_f=\langle xyz|f\rangle$, as shown in Fig.~\ref{fig5}(c)~\cite{11PRL_Ishida}. Then, $\int dz\,\psi_f^*(x,y,z)\partial_z\psi_i(x,y,z)\sim\psi_f^*(x,y,0)\int_{-a}^adz\,\partial_z\psi_i(x,y,z)=0$, and hence, condition~\eqref{cond1} can hold and the vertical node may emerge. When $|i\rangle$ is a two-component spinor, the condition is read for both the up- and down-spin components. 

In the CD patterns for the surface-related states shown in Fig.~\ref{fig3}(c), there is no apparent vertical node at $\theta_x=0^{\circ}$. This implies that condition~\eqref{cond1} is not fulfilled, or that the surface states observed in the ARPES image are not much localized in the surface region compared to $\lambda_{\rm{dB}}$. $\lambda_{\rm{dB}}$ can be estimated from the kinetic energy of the photoelectron $\varepsilon_k$: $\lambda_{\rm{dB}}\sim h/\sqrt{2m\varepsilon_k}$. Here, $h$ and $m$ are the Planck constant and electron mass, respectively. By using the relationship $\varepsilon_k=h\nu-w+(E-E_F)$ and setting the work function $w\sim4$~eV and $h\nu=10.8$~eV, $\lambda_{\rm{dB}}$ is estimated to be at most 5~\AA\ for the photoelectrons directly generated from the initial states at $E-E_F \ge$ -0.7~eV. According to the theoretical calculations~\cite{06PRL_Hirahara,16JPhysCondMat_Ishida}, the Bi(111) surface-state wavefunctions could penetrate into bulk for more than 5 bismuth bilayers, or $\gtrsim$20~\AA~\cite{05PRB_Bi111_LEED}. The absence of the vertical node thus supports these estimations. The deep penetration of the surface states can be attributed to their interaction with the bulk states~\cite{14PRB_Hsu}. Alternatively, if $\lambda_{\rm{dB}}$ can be sufficiently elongated by lowering $h\nu$, then the vertical node may emerge. A vertical-node-like pattern occurs around $E-E_F=-0.23$~eV [Fig.~\ref{fig3}(c)], but this is in the bulk-band region [see, the right panel of Fig.~\ref{fig3}(b)], and therefore, the node around -0.23~eV is understood as an accidental one, or beyond the classification scheme presented herein. 

The argument for neglecting $A_z$ through condition~\eqref{cond1} is similar to that adopted when explaining the vertical node that occurred in the CD patterns for intercalated~\cite{11PRL_Ishida} and aged~\cite{17PRB_Kondo} topological insulators. In those studies, $h\nu$ was as low as 7~eV and the confinement of the surface states could be enhanced by the intercalation~\cite{13PRB_CuBiSe_2D_LahoudKanigel,12NJP_Eremeev} and aging~\cite{10NCom_Hofmann_2DEG}. However, the samples had to be rotated step by step during the data acquisition because a slit-type analyzer was used instead of a slit-less analyzer. Therefore, the group theoretical argument was not rigorously applicable in those studies~\cite{11PRL_Ishida,17PRB_Kondo}.

The second case can occur when the incidence angle is $\eta=90^{\circ}$, namely when the circularly polarized photon field rotates on the $yz$ plane. The illustration $\varPsi_L$ [Fig.~\ref{fig5}(d)] operated on by either $\hat{\sigma}_y$ or $\hat{C}_2$ overlaps to the RCP illustration $\varPsi_R$ [Fig.~\ref{fig5}(e)], provided that the change in the direction of the light can be neglected, and that is valid when the long-wave-length approximation holds. Then, $\{\varPsi_L+\varPsi_R\}$ ($\{\varPsi_L-\varPsi_R\}$) with $\eta=90^{\circ}$ forms the basis for the irreducible representation $A_1$ ($B_2$) of ${\rm C}_{2\rm{v}}$; see Table~\ref{tab2}. The near-grazing-incidence configuration $\eta\sim 90^{\circ}$ can be achieved, for example, in the so-called Takata setup~\cite{05Nucl_Takata}, wherein the electron-lens axis of the analyzer is placed perpendicular to the incident hard-X-ray beam that can be circularly polarized~\cite{19SciTech_Ueda_HAXPES}. The Takata setup was the key to attain the throughput~\cite{07Takata} high enough for conducting ARPES even in the hard-X-ray regime~\cite{11NMat_HXARPES_Gray}.

\subsection{The case for crystalline samples}\label{s4c}
So far, we have investigated CD ARPES when the sample-and-detector $\varphi$ has $\rm{C}_{\rm{\infty v}}$ symmetry. The results evidenced that the symmetry of the experimental setup including the light is imprinted in the CD pattern. Thereby, we classified the CD pattern by using the group theory. We identified three types characterized by the point groups: $\rm{C}_{\sigma}$ (Section~\ref{s4a}), $\rm{C}_{2\rm{v}}$ (Section~\ref{s4b}), and $\rm{C}_{\infty\rm{v}}$ (Appendix~\ref{A1}).

Here, we extend the argument to the case when the sample is a crystal so that the symmetry of $\varphi$ is lower than $\rm{C}_{\rm{\infty v}}$. Four types will be identified as described below.

First, when the crystal has a mirror plane and that is matched to the plane of incidence ($y=0$), the arguments for $\rm{C}_{\sigma}$ presented in Section~\ref{s4a} can readily be applied; $I_L-I_R$ will have $A_2$ symmetry of $\rm{C}_{\sigma}$ and acquires the horizontal node. Second, when the crystal has another mirror plane at $x=0$ besides that at $y=0$, the arguments for $\rm{C}_{2\rm{v}}$ presented in Section~\ref{s4b} become applicable; when $A_z$ can be neglected from the light-electron interaction (when $\eta=90^{\circ}$), $I_L-I_R$ will have $A_2$ ($B_2$) symmetry of $\rm{C}_{2\rm{v}}$. Third, when the crystal has a mirror plane matched to $x=0$ but not at $y=0$, the arguments for $\rm{C}_{2\rm{v}}'=\{$$\hat{E},$ $\hat{\sigma}_x\}$ can be applied; when $A_z$ can be neglected from the light-electron interaction, $I_L-I_R$ will have $A_2$ symmetry of $\rm{C}_{2\rm{v}}'$. Finally, when the sample does not have a mirror plane at $y=0$, the two setups described by $\varPsi_L$ and $\varPsi_R$ cannot be converted to one other by any geometrical symmetry operations, and each of $\varPsi_L$ and $\varPsi_R$ is at most the base for the $A_1$ representation of the most primitive point group $\rm{C}_{1}=$ $\{$$\hat{E}$$\}$. 

To summarize, when the sample is a single crystal, we identify four types in the CD pattern characterized by the point groups $\rm{C}_{\sigma}$, $\rm{C}_{2\rm{v}}$, $\rm{C}_{\sigma}'$ and $\rm{C}_{1}$.

\subsection{Elliptical dichroism}\label{s4d}

The group theoretical classification of $I_L-I_R$ owes to the fact that RCP is the reflection of LCP with respect to the incidence plane $y=0$. In other words, the pair $\{{\bm A}_{L},{\bm A}_{R}\}$ being invariant with $\hat{\sigma}_y$, or having the mirror symmetry in short, was the essential ingredient for the arguments to hold.  Thus, as illustrated in the bottom row of Fig.~\ref{fig1}, there is no need for the pair to be composed of LCP and RCP. As we shall explicate below, most of the arguments can be retained even when the pair is of left- and right-elliptical polarizations (LEP and REP) as long as the pair has the mirror symmetry with respect to $y=0$.

\begin{figure}[htb]
\begin{center}
\includegraphics{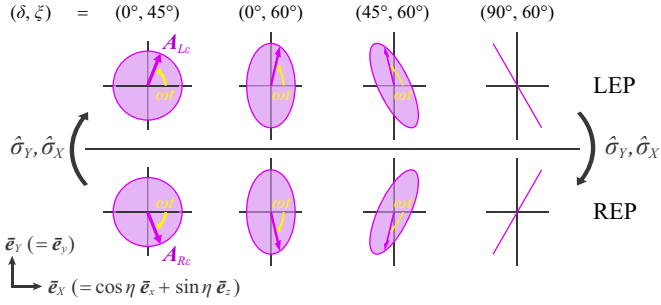}
\caption{\label{fig6}The mirror-symmetric pair of the polarized light. REP (bottom) is constructed by reflecting LEP (top) about $Y=0$. From left to right, the polarization changes from circular to linear as the parameter set $(\delta,\xi)$ is varied. Note, LEP and REP are also symmetric with respect to $X=0$.}
\end{center}
\end{figure}

To begin with, we set the pair of LEP and REP as follows. We first regard a particular polarized light directed along $\bar{\bm{e}}_Z$ as LEP, which can be described as the superposition of two orthogonal transverse waves as
\begin{eqnarray}
{\bm A}_{L\varepsilon}	&=&		A[\cos\xi\cos(\omega t +\delta)\bar{\bm{e}}_X + \sin\xi\sin(\omega t)\bar{\bm{e}}_Y] \nonumber\\ 
										&=&		{\bm A}_{+\varepsilon}e^{-i\omega t} + c.\,c., \nonumber
\end{eqnarray}
and then regard its mirror reflection with respect to the incidence plane $y=Y=0$ as REP:
\begin{equation}
{\bm A}_{R\varepsilon}\equiv\hat{\sigma}_y {\bm A}_{L\varepsilon}={\bm A}_{-\varepsilon}e^{-i\omega t}+c.\,c. \nonumber
\end{equation}
Here, $A\cos\xi$ ($A\sin\xi$) is the amplitude of the transverse wave polarized along $\bar{\bm{e}}_X$ ($\bar{\bm{e}}_Y$), $\delta$ sets the phase difference between the two waves, and $\bm{A}_{\pm\varepsilon}=A[e^{-i\delta}\cos\xi(\cos\eta\bar{\bm{e}}_x+\sin\eta\bar{\bm{e}}_z)\pm i\sin\xi\bar{\bm{e}}_y]/2$. Some pairs of LEP and REP are shown in Fig.~\ref{fig6}, in which the cases for linear and circular polarizations are included. For the moment, we exclude the special cases for the linear polarizations along $\bar{\bm{e}}_X$ (${\bm A}_X$) and $\bar{\bm{e}}_Y$ (${\bm A}_Y$), which can respectively be obtained by setting $(\delta,\xi)$ to $(90^{\circ}, 0^{\circ})$ and $(90^{\circ}, 90^{\circ})$; see, Section~\ref{s4e}. By definition, $\{{\bm A}_{L\varepsilon},{\bm A}_{R\varepsilon}\}$ becomes symmetric with respect to $\hat{\sigma}_y$, and the arguments for $\rm{C}_{\sigma}$ and $\rm{C}_{2\rm{v}}$ related to Figs.~\ref{fig4}, \ref{fig5}(d) and \ref{fig5}(e) are retained. Thus, the horizontal node seen in the CD pattern of Bi(111)/HOPG (Fig.~\ref{fig3}) ensures that the polarized pair had the mirror symmetry with respect to $y=0$.

The mirror symmetric pair of LEP and REP about the $y=Y=0$ plane thus defined automatically fulfills the mirror symmetry about the $X=0$ plane because the following relationship holds: ${\bm A}_{R\varepsilon}=-\hat{\sigma}_X{\bm A}_{L\varepsilon}$. Hence, the $A_z$-omitted pair $\{{\bm A}_{L\varepsilon}',{\bm A}_{R\varepsilon}'\}$ also becomes symmetric with respect to $\hat{\sigma}_x$, where ${\bm A}_{L\varepsilon,R\varepsilon}'={\bm A}_{L\varepsilon,R\varepsilon}-({\bm A}_{L\varepsilon,R\varepsilon}\cdot\bar{\bm{e}}_z)\bar{\bm{e}}_z$. Thus, the arguments for $\rm{C}_{2\rm{v}}$ related to Figs.~\ref{fig5}(a) and \ref{fig5}(b) are also retained.

With the change of the pair from circular to elliptical polarizations, the only case that is modified in the classification is when the light is incident in the surface normal; $\eta=0^{\circ}$ (Appendix~\ref{A1}). At $\eta=0^{\circ}$, $XYZ$ and $xyz$ coordinates are matched and the light-sample-detector entity becomes mirror symmetric with respect to both $x=0$ and $y=0$ planes. Thus, the elliptic dichroism pattern $I_{L\varepsilon}-I_{R\varepsilon}$ at $\eta=0^{\circ}$ will have the $A_2$ symmetry of $\rm{C}_{2\rm{v}}$ and acquire nodes along both $x=0$ and $y=0$.

\subsection{Linear vertical/horizontal polarization}\label{s4e}

The linear vertical ${\bm A}_Y$ and linear horizontal ${\bm A}_X$ polarizations were excluded from the elliptic dichroism arguments presented in the previous Section~\ref{s4d}. Both ${\bm A}_X$ and  ${\bm A}_Y$ remain themselves when operated on by $\hat{\sigma}_Y$ as well as by $\hat{\sigma}_X$. Thus, $\varPsi_{X, Y}\equiv{\bm A}_{X, Y}\oplus\varphi\oplus I_{X, Y}$ becomes a one-dimensional base for the most primitive representation (all indices in the character table are one) of the corresponding symmetry group of the experimental setup including the configuration of the light. As a result, the photoelectron distribution $I_{X, Y}$ itself will have the primitive symmetry with the reminder that $I_{X, Y}$ is understood as the subset of the base $\varPsi_{X, Y}$. 

Specifically, when the sample-and-detector has the $\rm{C}_{\infty{\rm v}}$ symmetry, $I_{X, Y}$ will be mirror symmetric with respect to $y=0$ ($A_1$ symmetry of  $\rm{C}_{\sigma}$) [Fig.~\ref{fig4}(a)]; if the conditions for the $\rm{C}_{2{\rm v}}$ symmetry hold, then $I_{X, Y}$ also will be mirror symmetric with respect to $x=0$ ($A_1$ symmetry of  $\rm{C}_{2{\rm v}}$) [Fig.~\ref{fig5}(a)]; when $\eta=90^{\circ}$ [Fig.~\ref{fig5}(d)] and the condition holds so that the direction of the light can be ignored, $I_Y$ will be symmetric with respect to both $y=0$ and $x=0$ ($A_1$ symmetry of  $\rm{C}_{2{\rm v}}$), whereas $I_X$ will be isotropic ($D_0^+$ symmetry of $\rm{C}_{\infty{\rm v}}$) because ${\bm A}_X$ has the infinite rotational and mirror symmetry of $\rm{C}_{\infty{\rm v}}$; and when $\eta=0^{\circ}$ [Fig.~\ref{fig_infty}(a)], $I_{X, Y}$ will have the $A_1$ symmetry of  $\rm{C}_{2{\rm v}}$. 

The classification for the cases when the sample is a crystal can also be obtained systematically. For example, when the crystal has a mirror plane at $y=0$, then $I_{X, Y}$ will have a mirror-symmetric distribution with respect to $y=0$ ($A_1$ symmetry of  $\rm{C}_{\sigma}$). Note, the classification scheme presented herein is different from the symmetry arguments for the selection rule that is often used to identify the orbital character of the bands; for example, see Ref.~\cite{12Sci_Okazaki}. Those arguments apply to the probability amplitude $\langle f|\bm{A}\cdot\hat{\bm{p}}|i\rangle$ for the photoelectrons emitted into the mirror plane of the crystal, and whether the amplitude is zero or not depends on whether $\bm{A}\cdot\hat{\bm{p}}|i\rangle$ can be regarded as even (allowed) or odd (forbidden) with respect to the mirror operation that keeps the crystal invariant.

\section{Summary and Remarks}\label{s5}

The motivation of the present study was to clarify how the direction of light can affect the dichroism seen in ARPES. To this end, we prepared a $\rm{C}_{\infty{\rm v}}$-symmetric sample, illuminated the sample with polarized light delivered from a laser-based source, recorded the distribution of the photoelectrons emitted from the sample by using a slit-less ARToF analyzer, and investigated whether the dichroism pattern seen in the distribution retained the $\rm{C}_{\infty{\rm v}}$ symmetry or not. The dichroism pattern was reduced from the $\rm{C}_{\infty{\rm v}}$ symmetry and exhibited a node along the plane of incidence set by the direction of the light. Thereby, we applied group theory and systematically classified the dichroism pattern with the direction of the light taken into account. 

The group-theoretical classification of the dichroism pattern described in the present study does not depend on the microscopic mechanism of the light-electron interaction, but with two exceptions: The long-wavelength approximation $\nabla\cdot\bm{A}=0$ was applied before condition~\eqref{cond1} and also when justifying the arguments in the experimental configuration of $\eta=90^{\circ}$. $\nabla\cdot\bm{A}$ can be non-negligible around the surface region when the dielectric response of the dipolar surface region ($\sim$5~\AA\, as estimated in a jellium model~\cite{79PRL_LevinsonFeibelman}) becomes substantial and modifies the spatial profile of the photon field at the length scale much shorter than the wavelength in vacuum. The dielectric response around the surface region can be substantial when the photon energy is below the valence (volume) plasmon energy existing at $\sim$10\,-\,30~eV~\cite{79PRL_LevinsonFeibelman,76SurfSci_Pendry,Kittel}, and can profoundly modify the matrix element for the very-surface-localized states~\cite{12PRL_Bian_BiAg_DivA,96PRL_MillerChiang}. Alternatively, the fingerprint of the surface photoelectric effect may manifest as the invalidation of the classification presented herein. 

We also remark that time-reversal operation $\hat{T}$ is not used in the classification. In fact, the event of a photoelectron emission cannot be symmetric with respect to $\hat{T}$. The photoelectron final state $|f\rangle$ has a time-reversal partner $\hat{T}|f\rangle$; $|f\rangle$ and $\hat{T}|f\rangle$ are the inverse LEED and LEED states, respectively, where LEED stands for low-energy electron diffraction. The two are degenerate in energy but are orthogonal to each other: $\langle f|\hat{T}|f\rangle=0$; see Section~29 of Ref.~\cite{68Schiff}. Thus, even when the hamiltonian of the system and its $N$-body initial state may possess $\hat{T}$-symmetry, the photo-excited $(N-1) +$ photoelectron state does not. 

As a final remark, the classification scheme with the direction of the light taken into accout can be understood as the generalization of the way introduced in Section~\ref{s1} to evaluate the existence of the dichroism: In Fig.~\ref{fig1}, the response of the sample to the polarized light was read as a value, namely, the amplitude of the current $i$, whereas the response detected in ARPES is the distribution in space. Thus, the scheme presented herein can be applied to classify the patterns in the distribution outputted from particle-in particle-out experimental setups, in which the incoming particle field is polarized.

\section*{Acknowledgments}
This work was conducted under the ISSP-CCES Collaborative Program and was supported by the Institute for Basic Science in Republic of Korea (IBS-R009-Y2 and IBS-R009-G2) and by JSPS KAKENHI (17K18749, 19K22140 and 19KK0350). Y.I.\ acknowledges the financial support by the University of Tokyo for the sabbatical stay at Seoul National University. S.S.\ acknowledges support from the Yonsei University BK21 program.

\appendix

\section{\label{A1}Representations of \textbf{C}$_{\infty\textbf{v}}$}
Here, we first summarize the irreducible representations of the group $\rm{C}_{\rm{\infty v}}$ after Section~4 of Ref.~[\onlinecite{76GroupT}]. Then, we apply the operators of ${\rm C}_{\rm{\infty v}}$ to an experimental setup which has a very-high symmetry, and show that CD under the symmetrical setup can be related to the base for the representation $D_0^-$ of  $\rm{C}_{\rm{\infty v}}$.

The group ${\rm C}_{\rm{\infty v}}$ consists of the rotation operation about the principal axis of angle $\alpha$, $\hat{R}(\alpha)$, and the mirror operation about any plane containing the principal axis.  All elements of ${\rm C}_{\rm{\infty v}}$ can be generated from  $\hat{R}(\alpha)$ and $\hat{\sigma}_y$, which are related to each other by $\hat{R}(\alpha)\hat{R}(\alpha')=\hat{R}(\alpha+\alpha')$ and $\hat{R}(\alpha)\hat{\sigma}_y=\hat{\sigma}_y \hat{R}(-\alpha)$.

Let us first consider the group ${\rm C}_{\rm{\infty}}=\{\hat{R}(\alpha)\}$, which is an invariant subgroup of ${\rm C}_{\rm{\infty v}}$: ${\rm C}_{\rm{\infty v}}={\rm C}_{\rm{\infty}} \oplus \hat{\sigma}_y{\rm C}_{\rm{\infty}}$. The base for the irreducible representation of ${\rm C}_{\rm{\infty}}$ can be $v_m$, which has the following property, $\hat{R}(\alpha) v_m=e^{-i\lambda\alpha} v_m$, where $m$ is an integer.

\begin{figure}[htb]
\begin{center}
\includegraphics{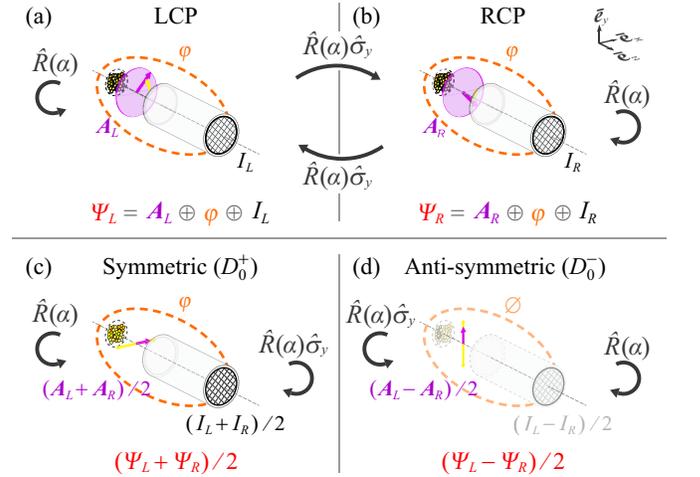}
\caption{\label{fig_infty} $\rm{C}_{\rm{\infty}\rm{v}}$ symmetry operations on the experimental setup. (a) Experimental setup $\varPsi_L$, in which LCP is in normal incidence to the sample.  (b) The only new illustration $\varPsi_R$ which can be constructed from $\varPsi_L$ by applying the operations of the group $\rm{C}_{\rm{\infty}\rm{v}}$. $\varPsi_R$ is identical to the setup where RCP is in normal incidence to the sample. (c and d) The illustrations that form the bases for the irreducible representations $D_0^+$ (c) and $D_0^-$ (d) of $\rm{C}_{\rm{\infty}\rm{v}}$.}
\end{center}
\end{figure}

Now, it is easy to see that $\{v_\lambda, v_{-\lambda}\}$ ($\lambda=$ 1, 2, 3, \ldots) forms the base for the irreducible representations of ${\rm C}_{\rm{\infty v}}$:
\begin{eqnarray}
\hat{R}(\alpha)\{v_\lambda, v_{-\lambda}\}			&=&		\{v_\lambda, v_{-\lambda}\}
\left[
\begin{array}{cc}
e^{-i\lambda\alpha} & 0   \\
0 & e^{i\lambda\alpha}
\end{array}
\right], \nonumber \\
\hat{\sigma}_y \{v_\lambda, v_{-\lambda}\}			&=&		\{v_\lambda, v_{-\lambda}\}
\left[
\begin{array}{cc}
0 & 1 \\
1 & 0
\end{array}
\right]. \nonumber
\end{eqnarray}
There are also two one-dimensional irreducible representations based by $v_0^\pm=v_\lambda\pm v_{-\lambda}$ ($\lambda=$ 0):
\begin{eqnarray}
\hat{R}(\alpha)v_0^\pm		&=&		v_0^\pm,		\label{eq_infty_1}	\\
\hat{\sigma}_y v_0^\pm		&=&		\pm v_0^\pm.		\label{eq_infty_2}
\end{eqnarray}

To summarize, for the ${\rm C}_{\infty{\rm v}}$, there are two one-dimensional irreducible representations $D_0^+$ and $D_0^-$ and a sequence of two-dimensional irreducible representations $D_\lambda$ ($\lambda=$ 1, 2, 3, \ldots).

Now, let us consider a highly-symmetric setup $\varPsi_L$, in which the LCP light is in normal incidence to the sample surface; see Fig.~\ref{fig_infty}(a). Such a setup cannot be realized in the present ARToF system because the analyzer blocks the light, but can be when a port is utilized to let light through slit-less-type analyzers such as the display-type analyzers~\cite{88RSI_Daimon}, momentum microscopes~\cite{16UltraMic_Tusche,20JJAP_Matsui} and hemispherical analyzers equipped with electron deflectors~\cite{18RSI_Ishida_Slitless}. In fact, the normal-incidence configuration was demonstrated and linear-polarization dependence was studied for $1T$-TaS$_2$~\cite{97PRB_Matsushita_TaS2} by using the display-type analyzer~\cite{93JJAP_Daimon}.

We apply all the operators of ${\rm C}_{\infty{\rm v}}$ to $\varPsi_L$. The only new illustration constructed through this procedure is $\varPsi_R$, which is displayed in Fig.~\ref{fig_infty}(b), and that happens to be identical to the setup where RCP is in normal incidence to the sample surface. Thus, $I_R$ that constitute the illustration $\varPsi_R$ is identified to the photoelectron distribution obtained with RCP.

The set $\{\varPsi_L, \varPsi_R\}$ becomes a base for the representation of ${\rm C}_{\infty{\rm v}}$, which is still reducible. It is $\{\varPsi_L\pm\varPsi_R\}$ that  becomes the base for the one-dimensional irreducible representation of ${\rm C}_{\infty{\rm v}}$; see Figs.~\ref{fig_infty}(c) and \ref{fig_infty}(d). Through the comparison to the one-dimensional representations $D_0^\pm$ expressed in eqs.~(\ref{eq_infty_1}) and (\ref{eq_infty_2}), $\varPsi_L - \varPsi_R$ is identified to the base for $D_0^-$, and $I_L-I_R=0$ is the subset of this base. Note, while $I_L-I_R$ is null, $\varPsi_L-\varPsi_R$ is not a null function because the subset of $\varPsi_L-\varPsi_R$ corresponding to the illustration of light, namely ${\bm A}_{L}-{\bm A}_{R}$ [Fig.~\ref{fig_infty}(d)], is not null.


\begin{thebibliography}{100}
\bibitem{footnote1}
 $\varphi (x, y, z)$ is a base for the $D_0^+$ representation of ${\rm
  C}_{\infty{\rm v}}$ (Appendix~A), because $\varphi$ remains itself when
  operated on by whatever rotation and reflection that consists ${\rm
  C}_{\infty{\rm v}}$.
\end{thebibliography}

\end{document}